\documentstyle [12pt]{article}
\def\H{{\cal H}}
\def\K{{\cal K}}

\newcommand{\lbl}[1]{\label{eq: #1}}

\def\R{{\rm I\hspace{-.15em}R}}

\def\Q{\mbox{l\hspace{-.47em}Q}}

\def\b{\begin{equation}}
\def\e{\end{equation}}
\def\bd{\begin{displaystyle}}
\def\ed{\end{displaystyle}}
\def\ba{\begin{array}}
\def\ea{\end{array}}

\def\bee{\begin{enumerate}}
\def\eee{\end{enumerate}}

\def\bes{\begin{eqnarray*}}
\def\ees{\end{eqnarray*}}
\def\be{\begin{eqnarray}}
\def\ee{\end{eqnarray}}

\begin{document}

\title{{\textbf{ Group theoretical interpretation of the\\ modified gravity in de Sitter space}}}
\author{M. Dehghani\thanks{e-mail:
m.dehghani@ilam.ac.com}} \maketitle \centerline{\it Department of
Physics, Ilam University, Ilam, Iran.}

\begin{abstract}

A framework has been presented for theoretical interpretation of
various modified gravitational models which is based on the group
theoretical approach and unitary irreducible representations (UIR's)
of de Sitter (dS) group. In order to illustrate the application of
the proposed method, a model of modified gravity has been
investigated. The background field method has been utilized and the
linearized modified gravitational field equation has been obtained
in the 4-dimensional dS space-time as the background. The field
equation has been written as the eigne-value equation of the Casimir
operators of dS space using the flat 5-dimensional ambient space
notations. The Minkowskian correspondence of the theory has been
obtained by taking the zero curvature limit. It has been shown that
under some simple conditions, the linearized modified field equation
transforms according to two of the UIR's of dS group labeled by
$\Pi^\pm_{2,1}$ and $\Pi^\pm_{2,2}$ in the discrete series. It means
that the proposed modified gravitational theory can be a suitable
one to describe the quantum gravitational effects in its linear
approximation on dS space. The field equation has been solved and
the solution has been written as the multiplication of a symmetric
rank-2 polarization tensor and a massless scalar field using the
ambient space notations. Also the two-point function has been
calculated in the ambient space formalism. It is dS invariant and
free of any theoretical problems.

\end{abstract}

{\it Proposed PACS numbers}: 04.62.+v, 03.70+k, 11.10.Cd, 98.80.H,
98.80.Jk

\newpage

\setcounter{equation}{0}
\section{Introduction}

It is well known that there are many good reasons to consider the
Einstein general relativity as the best theory for the gravitational
interaction, but according to the recent cosmological observations
it seems that this theory may be incomplete. In addition to the well
known problems of the Einstein general relativity in explaining the
astrophysical phenomenology (i.e., the galactic rotation curves and
small scale structure formation), recent cosmological data indicate
an underlying cosmic acceleration of the universe which cannot be
recast in the framework of the Einstein general relativity.

It is for these reasons and some other issues such as cosmic
microwave background anisotropies \cite{sp}, large scale structure
formation \cite{te}, baryon oscillations \cite{ei} and weak lensing
\cite{ja} that in recent years many authors are interested to
generalize standard Einstein gravity. Among alternative proposed
models the so-called extended theory of gravitation and, in
particular, the gravity theories stem from nonlinear actions or
higher-order theories of gravity have provided interesting results
\cite{cap, sta, cap2, ma, no, no2}. These models are based on
gravitational actions which are non-linear in the Ricci curvature
$\cal{R}$ and/ or contain terms involving combinations of
derivatives of $\cal{R}$ \cite{ke, tey, mag}.

Recent astronomical observations of supernova and cosmic microwave
background \cite{[4]} indicate that the universe is accelerating and
can be well approximated by a world with a positive cosmological
constant. If the universe accelerates indefinitely, the standard
cosmology leads to an asymptotic dS universe. In addition, dS
space-time plays an important role in the inflationary scenario
where an exponentially expanding approximately dS space-time is
employed to solve a number of problems in standard cosmology.
Furthermore, the quantum field theory on dS space-time is also of
considerable interest.

Furthermore, the gravitational field in the linear approximation
behaves like a massless spin-2 particle which propagates on the
background space-time. Following the Wigner's theorem, a linear
gravitational field should transform according to the UIR's of the
symmetric group of the background space-time. In this paper, dS
space-time has been considered as the background. It has been shown
that the proposed generalized Einstein's theory, in its linear
approximation, can be associated with the UIR's of dS group.

The main goal of this work is to propose a theoretical framework for
validity interpretation of the modified gravity theories, from group
theoretical point of view, in dS space. The idea is that if a
proposed model of modified gravity corresponds to the UIR,s of dS
group it can be considered as a possible successful model.

The organization of this paper is based on the following order. In
section-2, a generalized Einstein-Hilbert gravitational action has
been introduced and corresponding linear generalized Einstein
gravitational field equation has been obtained in terms of the
intrinsic dS coordinates as the background. Details of derivations
have been given in appendices. Next, the linearized field equation
has been written in terms of the Casimir operators of dS group
making use of the five-dimensional ambient space formalism. The
physical sector of the theory has been obtained by imposing the
divergenceless and traceless conditions and the possible relations
between this field equation and the UIR's of dS group have been
investigated. By imposing a simple condition the conformally
invariant theory of gravity is reproduced. In section-3, we obtained
the solution to the conformally invariant field equation, using the
ambient space notations. The solution can be written as the
multiplication of a symmetric generalized polarization rank-2 tensor
and a massless minimally coupled scalar field in dS space. In
section-4, we have calculated the conformally invariant two-point
function, in terms of the massless minimally coupled scalar
two-point function, using the ambient space formalism. It is dS
invariant, symmetric and satisfies the traceless and divergenceless
conditions. The results are summarized and discussed in section-5.
Some useful mathematical relations and details of derivations of
equations have been given in the appendices.

\setcounter{equation}{0}
\section{The field equation}

The terms containing fourth order derivatives of the metric may be
constructed out by curvature invariants (other than the cosmological
constant), that is $${\cal R},\;\;\; {\cal R}^2,\;\;\; {\cal
R}_{ab}{\cal R}^{ab},\;\;\; {\cal R}_{abcd}{\cal R}^{abcd}.$$

Therefore, the gravitational action for the modified field equation
in the 4-dimensional dS space-time with the metric signature
$(-,+,+,+)$ can be written in the following general form $$
I=\frac{1}{16\pi G}\int d^4 x\sqrt{-g}\left[a_0({\cal{R}}-2\Lambda )
+a_1{\cal{R}}^2+a_2{\cal{R}}^{ab}{\cal{R}}_{ab}+a_3{\cal{R}}^{abcd}{\cal{R}}_{abcd}
\right],$$ where $\Lambda=3H^2$ is the positive cosmological
constant. ${\cal{R}}_{abcd}$ is the Riemann tensor, ${\cal{R}}_{ab}$
is the Ricci tensor and ${\cal{R}}=g^{ab}{\cal{R}}_{ab}$ is the
Ricci scalar of the space-time under consideration. $a_0$, $a_1$,
$a_2$ and $a_3$ are constant coefficients. The coefficients $a_1$,
$a_2$ and $a_3$ are positive with the dimension of
$(\mbox{Length})^2$.

Taking note the fact that the Gauss-Bonnet action $$\frac{1}{16\pi
G}\int d^4 x\sqrt{-g}\left(
{\cal{R}}^2-4{\cal{R}}^{ab}{\cal{R}}_{ab}+{\cal{R}}^{abcd}{\cal{R}}_{abcd}\right)$$
is a total divergence. Adding it to the action will not contribute
to the field equations and enable us to simplify the action somewhat
and rewrite it as \b I=\frac{1}{16\pi G}\int d^4
x\sqrt{-g}\left[a_0({\cal{R}}-2\Lambda ) +
a{\cal{R}}^2+b{\cal{R}}^{ab}{\cal{R}}_{ab} \right],\e with new
coefficients. Therefore, including an
${\cal{R}}^{abcd}{\cal{R}}_{abcd}$  term is equivalent to altering
the coefficients. The theory described by this action is referred to
as fourth-order gravity, since it leads to fourth order equations.
Numerous papers have been devoted to the study of fourth-order
gravity.

Varying the action (2.1) with respect to the metric tensor $g_{ab}$
the modified gravitational field equation is obtained as
(Appendix-B)\b a_0 \H_{ab}^{(0)}+a \H_{ab}^{(1)}+b
\H_{ab}^{(2)}=0,\e where $\H_{ab}^{(0)}=G_{ab}+\Lambda g_{ab}$ and $
G_{ab}={\cal{R}}_{ab}-\frac{1}{2}{\cal{R}}g_{ab}$ is the Einstein
tensor and   \b \H_{ab}^{(1)}=2{\cal{R}}{\cal{R}}_{ab}-2\nabla_a
\nabla_b {\cal{R}}-\frac{1}{2}g_{ab}({\cal{R}}^2-4\Box {\cal{R}}),\e
\b \H_{ab}^{(2)}=\Box {\cal{R}}_{ab}-\nabla_c
\nabla_a{\cal{R}}^c_b-\nabla_c
\nabla_b{\cal{R}}^c_a+2{\cal{R}}_{a}^c{\cal{R}}_{cb}-\frac{1}{2}g_{ab}({\cal{R}}^{cd}{\cal{R}}_{cd}-2\nabla_c
\nabla_d{\cal{R}}^{cd}).\e Making use of the relations $[\nabla_c
\;, \nabla_a]{\cal{R}}^c_b={\cal{R}}_{da}  {\cal{R}}^d_{b}
-{\cal{R}}_{\;bca}^d   {\cal{R}}_{d}^c $, $\nabla_c
{\cal{R}}^c_{b}=1/2\nabla_b {\cal{R}}$, $\nabla_c \nabla_d
{\cal{R}}^{cd}=1/2\Box{\cal{R}}$ and other symmetry properties of
the Riemann tensor \cite{no2}, it is easy to show that the field
equation (2.2) is agree with Eq.(2.3) of ref. \cite{fan} with
$\gamma=0$.

\subsection{Linear field equation in dS space}

In order to obtain the linearized form of the field equation (2.2),
one can use the background field method. That is
$g_{ab}=g_{ab}^{(BG)}+h_{ab},$ in which $g_{ab}^{(BG)}$ is the
background metric and $h_{ab}$ is its fluctuations. Indices are
raised and lowered by the background metric. We suppose that
$g_{ab}^{(BG)}=g_{ab}^{(ds)}\equiv \tilde{g}_{ab}$. So one can write
\b g_{ab}=\tilde{g}_{ab}+h_{ab}\;\;\;\;\;\;\;
\mbox{and}\;\;\;\;\;\;\;g^{ab}=\tilde{g}^{ab}-h^{ab}.\e

The metric $\tilde{g}_{ab}$ is a solution to Einstein's field
equation with the positive cosmological constant $\Lambda=3H^2$: \b
\tilde{R}_{ab}-\frac{1}{2}\tilde{R}\tilde{g}_{ab}+3H^2
\tilde{g}_{ab}=0. \e

Using the approximations given in Eq.(2.5), in Eq.(2.3), we have
(Appendix-C) \b \H_{ab}^{(0)}= \tilde{H}_{ab}^{(0)}+ H_{ab}^{(0)},\e
where $\tilde{H}_{ab}^{(0)}$ is the dS correspondent to
$\H_{ab}^{(0)}$ and \b
H_{ab}^{(0)}=\frac{1}{2}(\nabla_{a}\nabla^{c}h_{bc}+\nabla_{b}\nabla^{c}h_{ac}-\Box
 h_{ab}-\nabla_{a}\nabla_{b}h'+2H^2h_{ab}) +\frac{1}{2}\tilde{g}_{ab}(\Box
h'-\nabla_{c}\nabla_{d}h^{cd}+H^2h'),\e in which $h'=h_{a}^{a}$ is
the trace of $h_{ab}$ with respect to the background metric and
$\nabla^b$ is the background covariant derivative. It is easy to
show that (Appendix-D) \b \H_{ab}^{(1)}=
\tilde{H}^{(1)}_{ab}+H_{ab}^{(1)}\e where $\tilde{H}_{ab}^{(1)}$ is
the correspondent to $\H_{ab}^{(1)}$ in dS space and

$$H_{ab}^{(1)}=+12H^2\left(\nabla_{a}\nabla^{c}h_{bc}+\nabla_{b}\nabla^{c}h_{ac}-\Box
h_{ab}\right)-2\nabla_a\nabla_b\left(\nabla_c\nabla_d h^{cd}-\Box
h'+3H^2h'\right)+24H^4h_{ab}$$
\b-2\tilde{g}_{ab}\left(3H^2\nabla_{c}\nabla_{d}h^{cd}+3H^4h'-\Box
\nabla_{c}\nabla_{d}h^{cd}+\Box^2h'\right).\e It is easy to show
that (Appendix-E) \b \H_{ab}^{(2)}=
\tilde{H}^{(2)}_{ab}+H_{ab}^{(2)}\e where $\tilde{H}_{ab}^{(2)}$ is
the correspondent to $\H_{ab}^{(2)}$ in dS space and
$$H_{ab}^{(2)}=\frac{1}{2}\left[\Box \left( \nabla_a \nabla_ch^c_b+\nabla_b \nabla_ch^c_a \right)
-2H^2\Box h_{ab}-\Box^2h_{ab}+ \nabla_a \nabla_b\Box h'\right]$$
$$+2H^2\left( \nabla_a \nabla_ch^c_b+\nabla_b \nabla_ch^c_a\right) -\nabla_a \nabla_b \nabla_{c}\nabla_{d}h^{cd}
-3H^2 \nabla_a \nabla_b h'+4H^4 h_{ab}$$
\b+\frac{1}{2}\tilde{g}_{ab}\left(2H^2\nabla_{c}\nabla_{d}h^{cd}-2H^4h'+7H^2\Box
h'+\Box \nabla_{c}\nabla_{d}h^{cd} -\Box^2h'\right) .\e

Substituting Eqs.(2.8), (2.10) and (2.12) in Eq.(2.2), we have \b
a_0H_{ab}^{(0)}+aH_{ab}^{(1)}+bH_{ab}^{(2)}=0.\e Eq.(2.13) is the
linearized modified gravitational field equation in dS background,
which has been written in terms of the intrinsic coordinates $X_a$
of the 4-dimensional dS space-time. The linear field equation (2.13)
can be written in the following explicit form
$$-\frac{b}{2}\Box^2
h_{ab}-\left(\frac{a_0}{2}+12aH^2+bH^2\right)\Box
h_{ab}+H^2(a_0+24aH^2+4bH^2)h_{ab}$$
$$+\left( 2a+\frac{b}{2}\right)\nabla_a\nabla_b \Box h'-\left(
\frac{a_0}{2}+6aH^2+3bH^2\right)\nabla_a\nabla_b
h'-(2a+b)\nabla_a\nabla_b \nabla_{c}\nabla_{d}h^{cd} $$
$$+\left(\frac{a_0}{2}+12aH^2+2bH^2+\frac{b}{2}\Box\right)\left( \nabla_{a}\nabla^{c}h_{bc}+\nabla_{b}\nabla^{c}h_{ac} \right)
$$
$$+\frac{1}{2}\tilde{g}_{ab}\left[\left(-a_0-12aH^2-2bH^2\right)\nabla_{c}\nabla_{d}h^{cd}+\left(a_0-bH^2\right)\Box
h'\right.$$
 \b \left. +\left(a_0-12aH^2-2bH^2\right)H^2h'+\left(4a+b\right)
\left(\Box\nabla_{c}\nabla_{d}h^{cd}-\Box^2 h'\right)\right]=0.\e

 The Minkowskian correspondence of the theory can be obtained by
taking the zero curvature (i.e. $H \rightarrow 0$) of Eq.(2.14), it
is
$$-\frac{1}{2}\left[b\Box^2
h_{ab}+a_0\Box h_{ab}-\left( 4a+b\right)\partial_a\partial_b \Box
h'-\left(a_0+b\Box\right)\left(
\partial_{a}\partial^{c}h_{bc}+\partial_{b}\partial^{c}h_{ac}
\right) +a_0\partial_a\partial_b h'\right]$$
\b+\frac{1}{2}\eta_{ab}\left[a_0\left(\Box
h'-\partial_{c}\partial_{d}h^{cd}\right)+(4a+b)\left(\Box
\partial_{c}\partial_{d}h^{cd} -\Box^2h'\right)\right]
-(2a+b)\partial_a\partial_b
\partial_{c}\partial_{d}h^{cd}=0,\e where $\eta_{ab}$ is the metric and
$\Box=\eta_{ab}\partial^a\partial^b=\partial^a\partial_a$ is the
wave operator in the flat space.

In order to obtain the physical sector of the model, one must to
impose the physical conditions $\nabla_a h^{ab}=0=\nabla_bh^{ab}$
and $h'=0$. In this case following Takook et al \cite{tatafa} we
obtain\b \left[-\frac{b}{2}\Box^2
-\left(\frac{a_0}{2}+12aH^2+bH^2\right)\Box
+H^2(a_0+24aH^2+4bH^2)\right]h_{ab}=0 ,\e fore the metric signature
$(-, +, +, +)$, and \b \left[-\frac{b}{2}\Box^2
+\left(\frac{a_0}{2}+12aH^2+bH^2\right)\Box
+H^2(a_0+24aH^2+4bH^2)\right]h_{ab}=0 ,\e fore the metric signature
$(+, -, -, -)$.

In the following subsection, in order to consider the possible
relations between the field equation and the UIR's of the dS group,
the linearized field equation (2.17) will be written in terms of the
Casimir operators of dS group, using the 5-dimensional ambient space
notations.

\subsection {dS group and Casimir operators in the field equation}
The dS space-time is a maximally symmetric space-time having a
positive constant curvature. It can be easily represented  as a
four-dimensional hyperboloid \b \eta_{\alpha\beta} x^\alpha
 x^\beta =-H^{-2},\;\;\; \alpha,\beta,...=0,1,2,3,4 ,\e embedded in
 a  flat five-dimensional space with metric $\eta_{\alpha\beta}=$diag$(1,-1,-1,-1,-1)$. The dS metrics is
\b ds^2=\eta_{\alpha\beta}dx^{\alpha}dx^{\beta}|_{x^2=-H^{-2}}=
\tilde{g}_{ab}dX^{a}dX^{b},\;\; a,b,...=0,1,2,3,\e where $X^a$'s are
the 4 space-time intrinsic coordinates in dS hyperboloid. Different
coordinate systems can be chosen \cite{mo}. Any geometrical object
in this space can be written in terms of the four local intrinsic
coordinates $X^a$ or in terms of the five global ambient space
coordinates $x^\alpha$.

In order to express Eq.($2.17$) in terms of the ambient space
notations, originally developed by Christian Fronsdal \cite{fr}, we
adopt the tensor field $\K_{\alpha\beta}(x)$ in ambient space
notations. Note that the ``intrinsic'' field $h_{ab}(X)$ is locally
determined by the transverse tensor field $\K_{\alpha\beta}(x)$
through \b\lbl{passage} h_{ab}(X)=\frac{\partial
x^{\alpha}}{\partial X^{a}}\frac{\partial x^{\beta}}{\partial
X^{b}}\K_{\alpha\beta}(x(X)). \e  In these notations, the solutions
to the field equations are easily written out in terms of scalar
fields. The reader how is not familiar to the ambient space
notations is referred to \cite{derota} and references therein.  The
symmetric tensor field $\K_{\alpha\beta}(x)$ is defined on dS
space-time and satisfies the transversality condition \cite{di,
gagata} \b x\cdot\K(x)=0,\mbox{ \it i.e. }x^\alpha
\K_{\alpha\beta}(x)=0,\mbox{ and } x^\beta \K_{\alpha\beta}(x)=0 .
\e The covariant derivative in the ambient space notations is
\b\label{eq:cov} D_\beta T_{\alpha_1 ...\alpha_i ...\alpha_n}= \bar
\partial_\beta T_{\alpha_1 ...\alpha_i
...\alpha_n}-H^2\sum_{i=1}^nx_{\alpha_i}T_{\alpha_1 ...\beta
...\alpha_n},\e where $\bar
\partial$ is  tangential (or transverse) derivative in
dS space \b \bar
\partial_\alpha=\theta_{\alpha \beta}\partial^\beta=
\partial_\alpha  +H^2x_\alpha x\cdot\partial,\;\;\;x\cdot\bar \partial=0,\e
$\theta_{\alpha \beta}=\eta_{\alpha \beta}+H^2x_{\alpha}x_{ \beta}$
is the transverse projector. It is easily shown that the metric
$\tilde{g}_{ab}$ corresponds to the transverse projector
$\theta_{\alpha\beta}$ that is \b \tilde{g}_{ab}(X)=\frac{\partial
x^{\alpha}}{\partial X^{a}}\frac{\partial x^{\beta}}{\partial
X^{b}}\theta_{\alpha\beta}(x). \e

The kinematical group of dS space is the $10$-parameter group
SO$_0(1,4)$ which is one of the two possible deformations of the
Poincar\'e group. There are two Casimir operators
\begin{equation}
Q^{(1)}_s=-\frac{1}{2}L_{\alpha\beta}L^{\alpha\beta},\qquad
Q^{(2)}_s=-W_{\alpha}W^{\alpha},\label{eq:cas}
\end{equation}
where
\begin{equation}
W_{\alpha}=-\frac{1}{8}\epsilon_
{\alpha\beta\gamma\delta\eta}L^{\beta\gamma}L^{\delta\eta},
\quad\mbox{with  10 infinitesimal generators}\quad
L_{\alpha\beta}=M_{\alpha\beta}+S_{\alpha\beta}.
\end{equation}
The subscript $s$ in $Q^{(1)}_s$, $Q^{(2)}_s$ reminds that the
carrier space is constituted by  tensors of rank $s$. The orbital
part $M_{\alpha\beta}$, and the action of the spinorial part
$S_{\alpha\beta}$ on a rank-2 tensor field ${\cal K}$ defined on the
ambient space read respectively \cite{gagata}
\begin{equation}
M_{\alpha\beta}=-i
(x_\alpha\partial_\beta-x_\beta\partial_\alpha),\qquad
S_{\alpha\beta}\K_{\gamma\delta}=-i
(\eta_{\alpha\gamma}\K_{\beta\delta}-\eta_{\beta\gamma}\K_{\alpha\delta}
+\eta_{\alpha\delta}\K_{\beta\gamma}-\eta_{\beta\delta}\K_{\alpha\gamma}).\e

The symbol $\epsilon_{\alpha\beta\gamma\delta\eta}$ holds for the
usual antisymmetrical tensor. The action of the Casimir operator
$Q_2^{(1)}$  on ${\cal K}$ can be written in the more explicit form
\begin{equation}\label{eq:act}
 Q_2^{(1)}{\cal
K}(x)=\left(Q_{0}^{(1)}-6\right){\cal K}(x)+2\eta {\cal K}'+2{\cal
S} x\partial\cdot{\cal K}(x)-2{\cal S}  \partial x\cdot{\cal K}(x),
\end{equation} where,
$Q_{0}^{(1)}=-{{1}\over{2}}M_{\alpha\beta}M^{\alpha\beta}=-H^{-2}(\bar\partial)^2$
is the scalar Casimir operator. The symmetrizer ${\cal S}$ is
defined for two vectors $\xi_{\alpha}$ and $\omega_{\beta}$ by
${\cal
S}(\xi_{\alpha}\omega_{\beta})=\xi_{\alpha}\omega_{\beta}+\xi_{\beta}\omega_{\alpha}$.
$\K'$ is the trace of the tensor $\K$ and the action of the Casimir
operator $Q_1^{(1)}$  on the vector $ K$ can be written in the more
explicit form
\begin{equation}\label{eq:act}
Q_1^{(1)} K(x)=\left(Q_{0}^{(1)}-2\right) K(x)+2x
\bar{\partial}\cdot K(x)+2H^2 x\;x\cdot K(x)-2  \bar{\partial}\;
x\cdot K(x).
\end{equation}

As shown by Dixmier\cite{ms5}, the UIR,s of dS group have a
classification scheme in terms of a pair of parameters $(p,q)$. The
Casimir operators take the following possible spectral values:
\begin{equation}
\langle Q^{(1)}_{p} \rangle=-p(p+1)-(q+1)(q-2) ,\qquad\quad \langle
Q^{(2)}_{p} \rangle=-p(p+1)q(q-1)\,.
\end{equation}
Depending on the different values of the pair of parameters $(p,q)$,
three different series of representations are distinguishable: the
principal, the complementary and the discrete series \cite{ms5,ms6}.
Mathematical details of the group contraction and the physical
principles underlying the relationship between dS and Poincar\'e
groups can be found in Refs \cite{ms7} and \cite{ms8} respectively.
The spin-$2$ tensor representations relevant to the present work are
\cite{derota}:
\begin{itemize}
\item[i)] The UIR's of the principal series labeled by $U^{2,\nu}$
with $p=s=2$ and $q=\frac{1}{2 }+i\nu$ correspond to the Casimir
spectral values:
\begin{equation}
\langle Q_2^{(1)}\rangle=\nu^2-\frac{15}{4},\;\;\nu \in \R,
\end{equation}
note that $U^{2,\nu}$ and $U^{2,-\nu}$ are equivalent. \item[ ii)]
The UIR's of the complementary series denoted by $V^{2,q}$ with
$p=s=2$ and $q-q^2=\mu,$ correspond to the following spectral values
\begin{equation}
\langle  Q_2^{(1)}\rangle=q-q^2-4\equiv
\mu-4,\;\;\;0<\mu<\frac{1}{4}\,.
\end{equation}
\item[iii)] The UIR's of the discrete series conventionally labeled by $\Pi^{\pm}_{2,q}$ in
which $p=s=2$ and takes the following spectral values
\begin{equation}
\langle Q_2^{(1)}\rangle=-6-(q+1)(q-2), \;\;q=1,2.
\end{equation}
\end{itemize}
The ``massless'' spin-2 field in dS space corresponds to the
$\Pi^{\pm}_{2,2}$ and $\Pi^{\pm}_{2,1}$ cases in which the sign
$\pm$, stands for the helicity. In these cases, the two
representations $\Pi^{\pm}_{2,2}$, in the discrete series with
$p=q=2$, have a Minkowskian interpretation. It is important to note
that the representations $\Pi^{\pm}_{2,1}$ do not have corresponding
flat limit \cite{derota}. (More details can be found in
\cite{gagata} and references therein.)

We now attempt to express the wave equation (2.17) in terms of the
Casimir operators of dS group. The d'Alembertian operator becomes
\cite{ta1}
\begin{equation}
\Box h_{ab}=\nabla^{c}\nabla_{c}h_{ab}=\frac{\partial
x^\alpha}{\partial X^a}\frac{\partial x^\beta}{\partial X^b}
\left[-H^2Q_{0}^{(1)}-2H^2\right]{\cal K}_{\alpha \beta} \;,
\end{equation} and
\begin{equation}
\Box^2
h_{ab}=\nabla^{c}\nabla_{c}\nabla^{d}\nabla_{d}h_{ab}=\frac{\partial
x^\alpha}{\partial X^a}\frac{\partial x^\beta}{\partial X^b}
\left[H^4\left(Q_{0}^{(1)}\right)^2+4H^4Q_{0}^{(1)}+4H^4\right]{\cal
K}_{\alpha \beta} \;,
\end{equation}

where, the conditions of tracelessness and divergence free (e.i.
$\bar{\partial}.\K=0=\K'$), have been imposed to the physical
states. By use of the above equations in Eq. (2.17) we have \b
\left[\frac{b}{2}H^2\left(Q_{0}^{(1)}\right)^2
+\left(\frac{a_0}{2}+12aH^2+3bH^2\right)Q_{0}^{(1)}
\right]\K_{\alpha\beta}=0. \e
 In terms of
different choice of coefficients in the proposed action (2.1)
different gravitational theories may be achieved. Now the following
various choices are considerable

$\bullet$ By choosing $a=0$ and $b=0$ we return to the physical
linear pure dS theory, that is
 \b Q_0^{(1)}\K_{\alpha\beta}=0, \;\;\;\;\;\; \mbox{or}\;\;\;\;\;\; \left(Q_2^{(1)}+6\right)\K_{\alpha\beta}=0.\e
This is an eigen-value equation with the eigen-value $\langle
Q_2^{(1)}\rangle=-6$. From the group theoretical point of view this
corresponds to UIR's of dS group labeled by $\Pi^{\pm}_{2,2}$ in the
discrete series which reduces to the physical representations of the
Poincar\'e group in the zero curvature limit. This is why it is
called as the physical state. It has been discussed in
\cite{derota}, for the gauge-fixed value equal to zero, \cite{de2}
for the gauge-fixed value equal to $\frac{2}{5}$ and the extended
discussions are given in \cite{de3}.

$\bullet$ Letting $a_0=1$, $b=0$, the model reduces to a
$f(\cal{R})$ theory model with $f({\cal{R}})={\cal{R}} +a
{\cal{R}}^2$. It is known as a relatively successful model, which
explains the inflation and positive acceleration of the universe
\cite{1, 2}. Under these conditions, the linearized field equation
(2.39) reduces to  \b
(1+24aH^2)Q_0^{(1)}\K_{\alpha\beta}=0,\;\;\;\;\;\;
\mbox{or}\;\;\;\;\;\;(1+24aH^2)\left(Q_2^{(1)}+6\right)\K_{\alpha\beta}=0.\e
It corresponds to the UIR's of dS group labeled by $\Pi^{\pm}_{2,2}$
in the discrete series too. This is why the model is a successful
one. The field equation (2.38) has been considered in ref.
\cite{deka}.

$\bullet$ One may set $a_0=0$, $a=-\frac{1}{3}$ and $b=1$, by which
the theory reduces to the Weyl conformal theory with the linearized
field equation \b
Q_0^{(1)}\left(Q_0^{(1)}-2\right)\K_{\alpha\beta}=0,\;\;\;\;\;\;
\mbox{or}\;\;\;\;\;\;
\left(Q_2^{(1)}+6\right)\left(Q_2^{(1)}+4\right)\K_{\alpha\beta}=0.\e
The same equation has been obtained by Dehghani, et. al from a
different approach in \cite{derota}.

The field equation $\left(Q_2^{(1)}+4\right)\K_{\alpha\beta}=0,$ is
also an eigen-value equation with the eigen-value $\langle
Q_2^{(1)}\rangle=-4$. It corresponds to one of the UIR's of dS group
denoted by $\Pi^{\pm}_{2,1}$ in the discrete series with the same
Poincar\'e correspondence as $\Pi^{\pm}_{2,2}$ in the zero curvature
limit. Indeed two of UIR's of dS group have only one Poincar\'e
correspondence. It has been discussed in \cite{de}.

As it is clear with the help of above-mentioned examples, we believe
that it is necessary for any successful theory of gravity to
transform according to the UIR,s of dS group. In other words if a
model of modified gravity theory does not correspond to the UIR,s of
dS group in its linear approximations it can not produce valid and
helpful physical results.

For the general discussion on the proposed modified gravity theory,
let $A=aH^2$ and $B=bH^2$. In terms of these dimensionless
coefficients the field equation (2.36) can be written as \b
\left[\left(Q_0^{(1)}\right)^2+\left(\frac{a_0}{B}+24\frac{A}{B}+6\right)Q_0^{(1)}\right]\K_{\alpha\beta}=0
,\;\;\; B\neq 0.\e As a direct mathematical result, the proposed
model in it's linear approximation, generally transforms according
to the UIR's of dS group and it is a suitable candidate model of
gravitation on dS space if the characteristic equation \b
\frac{a_0}{B}+24\frac{A}{B}+8=0 ,\;\;\; B\neq 0 ,\e is satisfied.
Under this condition it describes a massless spin-2 particle (the
graviton, if it exists) in it's linear approximation and transforms
according to two of UIR,s of dS group. We therefore believe that it
can be a successful modified gravity theory. For more clarity, in
the following sections, we solve the field equation ($2.40$), with
the condition (2.41), using the ambient space formalism. Also we
obtain the two-point function for the linearized theory of
gravitation making use of the ambient space notations, and show that
the results are free of any theoretical problems.

\setcounter{equation}{0}
\section{Solution to the conformal field equation}

A general solution of to the conformall field equation can be
constructed from the combination of a scalar field and two vector
fields. Let us first introduce a traceless and transverse tensor
field $\K$ in terms of a five-dimensional constant vector
$Z_1=(Z_{1\alpha})$ and a scalar field $\phi_1$ and two vector
fields $K$ and $K_g$ by putting \cite{gagata, derota, de2, de3, de,
tap} \b \K=\theta\phi_1+ {\cal {S}}\bar {Z}_1K+D_2K_g,\e where $D_2$
is the generalized gradient operator defined by $D_2K={\cal{
S}}(D_1+x)K$, $D_{1\alpha}=H^{-2}\bar{\partial}_\alpha$ and $\bar
Z_{1\alpha}=\theta_{\alpha\beta} Z^{\beta}_1$. Taking the trace of
$\K_{\alpha\beta}$ we have
\b\K'=4\phi_1+2Z_1.K+2H^2(x.Z_1)x.K+2D_1.K_g-2x.K_g=0,\e Using the
ansatz ($3.1$) to the field equation we have (Appendix F)  \b
\left\{\begin{array}{ll}
          (Q^{(1)}_0+4)(Q^{(1)}_0+6)\phi_1+8(Q^{(1)}_0+2)Z_1.K=0,\;\;\;\;\;\;\;\;\;\;\;\;\;\;\;\;\;\;\;\;\;\;\;\;\;\;\;$(a)$\\
          \\Q^{(1)}_1\left(Q^{(1)}_1+2\right)K=0,\;\;\;\;\mbox{or}\;\;\;\;Q^{(1)}_1Q^{(1)}_0K=0,\;\;\;\partial .K=0=x.K,\;\;\;\;\;\;\;\;\;$(b)$\\
          \\(Q^{(1)}_1+4)(Q^{(1)}_1+6)K_g =4H^2\left[ (Q^{(1)}_1+5)x.Z_1K+Z_1.D_1K -xZ_1.K\right].  \;\;\;$(c)$
        \end{array} \right.\e
The vector field $K$ can be written in the following general form \b
K_\alpha=\bar{Z}_{2\alpha}\phi_2+D_{1\alpha}\phi_3,\e where $Z_2$ is
another constant 5-vector and $\phi_2$ and $\phi_3$ are two
arbitrary scalar fields, should be determined. Using the
divergenceless condition we have \b Q^{(1)}_0
\phi_3=Z_2.\bar{\partial}\phi_2+4H^2(x.Z_2)\phi_2,\e and
substituting Eq.(3.4) in Eq.(3.3-b) leads to the following two
equations \b Q^{(1)}_0(Q^{(1)}_0-2)\phi_2=0,\e \b
Q^{(1)}_0(Q^{(1)}_0+2)\phi_3=4H^2Q^{(1)}_0[(x.Z_2)\phi_2]+8H^2(x.Z_2)\phi_2+4Z_2.\bar{\partial}\phi_2.\e
The Eq.(3.6) has a dS plane wave solution of the form \b\phi_2=(H
x.\xi)^\sigma, \;\;\;\;\;\;\xi^2=0,
\;\;\;\;\;\;\;\;\;\mbox{with}\;\;\;\;\;\;\;\;\sigma(\sigma+3)(\sigma+2)(\sigma+1)=0.\e
Note that $\phi_2$ is the minimally coupled scalar field for
$\sigma=0, -3$. In that case it obeys the field equation $Q^{(1)}_0
\phi_2=0 $ \cite{derota, gareta}. Also $\phi_2$ is the conformally
coupled scalar field for $\sigma=-1, -2$ and satisfies the field
equation $(Q^{(1)}_0-2) \phi_2=0 $ \cite{berota}.

Substituting  $Q^{(1)}_0 \phi_3$ and $(Q^{(1)}_0)^2 \phi_3$ from
Eq.(3.5) into Eq.(3.7), we obtain \b Q^{(1)}_0
Z_2.\bar{\partial}\phi_2=2Z_2.\bar{\partial}\phi_2.\e
 Now regarding Eqs.(3.6) and (3.9) and using the identity \b
Q^{(1)}_0[(x.Z_2)\phi_2]=(x.Z_2)Q^{(1)}_0\phi_2-4(x.Z_2)\phi_2-2Z_2.D_1\phi_2
,\e   we obtain \b
Q^{(1)}_0[(x.Z_2)Q^{(1)}_0\phi_2]=2Q^{(1)}_0[(x.Z_2)\phi_2]+8(x.Z_2)\phi_2+4Z_2.D_1\phi_2
.\e Combining Eqs.(3.9) and (3.11) we have \b
(x.Z_2)\phi_2=\frac{1}{8}Q^{(1)}_0\left[
(x.Z_2)Q^{(1)}_0\phi_2-2(x.Z_2)\phi_2-2Z_2.D_1\phi_2\right].\e
Substituting Eqs.(3.9) and (3.12) in Eq.(3.5) we have  \b
\phi_3=\frac{1}{2}\left[H^2(x.Z_2)Q^{(1)}_0\phi_2-Z_2.\bar{\partial}\phi_2-2H^2(x.Z_2)\phi_2\right].\e
Now the solution to Eq.(3.3-b) can be written in terms of the dS
massless scalar field $\phi_2\equiv \phi_s$  as \b
K_\alpha=\bar{Z}_{2\alpha}\phi_s+\frac{1}{2}D_{1\alpha}\left[H^2(x.Z_2)Q^{(1)}_0-Z_2.\bar{\partial}-2H^2x.Z_2\right]\phi_s.\e
The explicit form of the vector field $K_\alpha$ is \b
K=\frac{\sigma}{2}\left[(\sigma+2)\bar
Z_{2}+(\sigma^2+2\sigma-2)\frac{x.Z_2}{x.\xi}\bar\xi
\right]\phi_s,\e and the condition of $\bar{\partial}.K_\alpha=0$
can be written in the following explicit form \b
\bar{\partial}.K=\frac{1}{2}\sigma^2(\sigma+3)(\sigma+4)(x.Z_1)(x.Z_2)\phi_s=0.\e
Noting Eq.(3.8), it is valid only for $\sigma=0,-3.$ As pointed out
before we can treat the scalar field $\phi_s$ as the massless
minimally coupled scalar field. Furthermore under these
circumstances the vector field $K_\alpha$ satisfies the relation \b
Q^{(1)}_0 K_\alpha=0,\;\;\;\;\;\;\;\;\;\;\;\;
\mbox{or}\;\;\;\;\;\;\;\;\;\;\;\;\left(Q^{(1)}_1+2\right)K=0. \e

It is easy to show that Eq.(3.3-a) has a solution of the form
\b\phi_1=-\frac{2}{3}Z_1.K,\;\;\;\;\;\;\;\;\;
Q^{(1)}_0(Q^{(1)}_0-2)\phi_1=0.\e It means that $\phi_1$ satisfies
the scalar massless field equation in dS space \cite{gareta,
berota}.  Now Eq (3.2) can be written as \b
\bar{\partial}.K_g=\frac{1}{3}H^2Z_1.K,\;\;\;\;\;\;\;\;\;\;\;\;
x.K_g=0.\e  Making use of the relation
$$(Q^{(1)}_1+5)x.Z_1K=2x(Z_1.K)-2(Z_1.D_1) K- (x.Z_1)K,$$
Eq.(3.3-c) can be written as \b (Q^{(1)}_1+4)(Q^{(1)}_1+6)K_g
=4H^2\left[xZ_1.K-x.Z_1K-Z_1.D_1K \right].\e Now using the
identities $$6(x.Z_1)K=(Q^{(1)}_1+6)\left[(x.Z_1)K+\frac{1}{9}
D_1(Z_1.K)\right],$$ $$2(xZ_1.K-Z_1.D_1K) =(Q^{(1)}_1+6)(x.Z_1K),$$
in Eq.(3.20) we obtain \b (Q^{(1)}_1+4)K_g
=\frac{4}{3}H^2\left[(x.Z_1)K-\frac{1}{18}D_1(Z_1.K )\right].\e It
is easy to show that \b 4D_1(Z_1.K)=(Q^{(1)}_1+4)D_1(Z_1.K) ,\e \b
4(x.Z_1)K=(Q^{(1)}_1+4)\left[(x.Z_1)K+\frac{1}{6}D_1(Z_1.K)\right]
.\e Combining Eqs.(3.21)-(3.23) results in
 \b  K_g =\frac{1}{3}H^2\left[(x.Z_1)K+\frac{1}{9} D_1(Z_1.K)\right].\e
It satisfies the conditions given in Eq.(3.19).

Substituting Eqs.(3.15), (3.18) and (3.24) in Eq.(3.1) one can show
that \b \K_{\alpha \beta}(x)= {\cal
E}_{\alpha\beta}(x,\xi,Z_1,Z_2)\phi_s,\e where $\phi_s$ is a
massless scalar field in dS space and ${\cal E}$ is a generalized
symmetric polarization tensor, \b  {\cal E}=\frac{\sigma}{2}
\left[-\frac{2}{3}\theta Z_1. +{\cal S}\bar{Z}_1+H^2\frac{1}{3}D_2
(x.Z_1+\frac{1}{9}D_1Z_1.) \right]\left[(\sigma+2)\bar
Z_{2}+(\sigma^2+2\sigma-2)\frac{x.Z_2}{x.\xi}\bar\xi   \right].\e
 It is consistent with the results in \cite{derota} and \cite{de3} with $c=0$.

\setcounter{equation}{0}
\section{The conformal two-point function }

The two-point function ${\cal W}_{\alpha\beta \alpha'\beta'}(x,x')$,
which is a solution of the wave equation with respect to $x$ or
$x'$, can be found simply in terms of the scalar two-point function.
Very similar to the recurrence formula ($3.1$) let us try the
following possibility \cite{gagata, derota, de2, de3, de, tap}
 \b {\cal W}(x,x')=\theta \theta'{\cal W}_0(x,x')+{\cal
S}{\cal S}'\theta.\theta'{\cal W}_{1}(x,x')+D_2D'_2{\cal
W}_g(x,x'),\e where ${\cal W}$, ${\cal W}_{1}$ and ${\cal W}_{g}$
are transverse bi-vectors, ${\cal W}_{0}$ is bi-scalar and
$D_2D'_2=D'_2D_2$. Substituting the two-point function ($4.1$) in
the field equation with respect to $x$, we have \b \left\{
\begin{array}{ll}
          (Q^{(1)}_0+4)(Q^{(1)}_0+6)\theta'{\cal W}_0+8(Q^{(1)}_0+2){\cal S}'\theta'.{\cal W}_{1},\;\;\;\;\;\;\;\;\;\;\;\;\;\;\;$(a)$\\
          \\Q^{(1)}_1(Q^{(1)}_1+2){\cal W}_{1}=0, \;\;\;\;\;\;\mbox{or}\;\;\;\;\;\;Q^{(1)}_1Q^{(1)}_0{\cal W}_{1}=0,\;\;\;\;\;\partial.{\cal W}_{1}=0,\;\;\;\;\;\;\;\;\;$(b)$\\
          \\(Q^{(1)}_1+4)(Q^{(1)}_1+6)D'_2{\cal W}_g=4H^2{\cal S}'\left[(Q^{(1)}_1+5)(x.\theta'){\cal W}_{1}+\theta'.D_1{\cal W}_{1}+x\theta'.{\cal W}_{1}\right].\;\;\;$(c)$
        \end{array}\right.\e

The solution to Eq.(4.2-b) has the following general form \b {\cal
W}_{1}=\theta.\theta'{\cal W}_{2}+D_1D'_1{\cal
W}_{3},\;\;\;\;\;\;\;\;\; \mbox{and} \;\;\;\;\;\;\;\;\; D'_1{\cal
W}_{3}=\frac{1}{2}\left[H^2(x.\theta')Q^{(1)}_0-\theta'.\bar{\partial}-2H^2x.\theta'\right]{\cal
W}_{2},\e
 in which ${\cal
W}_{2}\equiv{\cal W}_{s}$ is the massless minimally coupled scalar
two-point function. The dS-invariance two-point function for the
massless minimally coupled scalar field in the ``Gupta-Bleuler
vacuum'' state is \cite{ta3} \b {\cal
W}_{s}(x,x')=\frac{iH^2}{8\pi^2} \epsilon (x^0-x'^0)[\delta(1-{\cal
Z}(x,x'))+\vartheta ({\cal Z}(x,x')-1)],
\e with \b {\cal{Z}}=-H^2x.x',\;\;\;\;\;\; \mbox{and} \;\;\;\;\;\; \epsilon (x^0-x'^0)=\left\{ \ba{rcl} 1&x^0>x'^0 ,\\
0&x^0=x'^0 ,\\ -1&x^0<x'^0.\\ \ea\right.\e

 In summary, the solution to the above system of equations is
\b {\cal
W}_{1}=\left[\theta.\theta'+\frac{1}{2}D_1\left(H^2x.\theta'Q^{(1)}_0-\theta'.\bar{\partial}-2H^2x.\theta'
\right)\right]{\cal W}_{s},\e
 \b \theta'{\cal
W}_0(x,x')=-\frac{2}{3}{\cal S}'\theta'.{\cal W}_{1}(x,x') ,\e  \b
D'_2{\cal W}_g(x,x')=\frac{1}{3}H^2 {\cal S'}\left[(x.\theta'){\cal
W}_{1}+\frac{1}{9}D_1(\theta'.{\cal W}_{1})\right].\e  The two-point
function ($4.1$) also satisfies the field equation with respect to
$x'$, in this case one can obtain\b \left\{
\begin{array}{ll}
          (Q'^{(1)}_0+4)(Q'^{(1)}_0+6)\theta{\cal W}_0+8(Q'^{(1)}_0+2){\cal S}\theta.{\cal W}_{1},\;\;\;\;\;\;\;\;\;\;\;\;\;\;\;$(a)$\\
          \\Q'^{(1)}_1(Q'^{(1)}_1+2){\cal W}_{1}=0, \;\;\;\;\;\;\mbox{or}\;\;\;\;\;\;Q'^{(1)}_1Q'^{(1)}_0{\cal W}_{1}=0,\;\;\;\;\;\partial'.{\cal W}_{1}=0,\;\;\;\;\;\;\;\;\;$(b)$\\
          \\(Q'^{(1)}_1+4)(Q'^{(1)}_1+6)D_2{\cal W}_g=4H^2{\cal S}\left[(Q'^{(1)}_1+5)(x'.\theta){\cal W}_{1}+\theta.D'_1{\cal W}_{1}-x'\theta.{\cal W}_{1}\right].\;\;\;$(c)$
\end{array}\right.\e
with the solutions
 \b {\cal
W}_{1}=\left[\theta'.\theta+\frac{1}{2}D'_1\left(H^2x'.\theta
Q'^{(1)}_0-\theta.\bar{\partial}'-2H^2x'.\theta \right)\right]{\cal
W}_{s},\e
 \b \theta{\cal W}_0(x,x')=-\frac{2}{3}{\cal
S}\theta.{\cal W}_{1}(x,x') ,\e  \b D_2{\cal
W}_g(x,x')=\frac{1}{3}H^2 {\cal S}\left[(x'.\theta){\cal
W}_{1}+\frac{1}{9}D'_1(\theta.{\cal W}_{1})\right].\e Note that the
primed operators act on the primed coordinates only.

 Making use of Eqs.($4.6$)-($4.8$) or ($4.10$)-($4.12$) one can show that the conformal two-point function can
 be written as
\b {\cal
W}_{\alpha\beta\alpha'\beta'}=\Delta_{\alpha\beta\alpha'\beta'}
{\cal W}_s ,\e where $$ \Delta=\frac{1}{6}\left[-2\theta {\cal
S}'\theta'.+{\cal S}{\cal S}'\theta.\theta'+H^2D_2{\cal
S}'(x.\theta'+\frac{1}{3}D_1\theta'.) \right]$$
\b\times\left[2\theta.\theta'+D_1\left(H^2x.\theta'
Q^{(1)}_0-\theta'.\bar{\partial}-2H^2x.\theta' \right)\right] .\e It
agrees with the results in \cite{derota} and \cite{de3} with $c=0$.

\section{Conclusion }
According to the recent cosmological observations it seems that the
standard Einstein theory of gravity may be incomplete and many
attempts have been made to modify this theory. The so-called
modified theory of gravitation and, in particular, non-linear
gravity theories or higher-order theories of gravity have provided
interesting results. The proposed models are based on gravitational
actions which are non-linear in the Ricci curvature and constructed
out by curvature invariants.

This work is devoted to an extension of the Einstein-Hilbert
gravitational action, which is constructed out by the linear
combination of Ricci scalar and Ricci tensor invariants in dS space.
Varying the proposed action with respect to metric tensor leads to a
fourth order gravitational field equation, conventionally named as
the modified gravitational theory. The background field method is
utilized and the linearized field equation is obtained in terms of
intrinsic coordinates in the 4-dimensional dS space as the
background.

The gravitational field in the linear approximation behaves like a
massless spin-2 particle which propagates on the background
space-time. According to Wigner's theorem, a linear gravitational
field should transform according to the UIR's of the symmetry group
of the background space-time. In order to investigate the possible
relations between the field equation and the UIR's of dS group it is
transformed into the flat five-dimensional ambient space and the
linearized field equation is written in terms of the Casimir
operators of dS group. We obtained the Minkowskian correspondence of
the theory by taking the zero curvature limit. The physical sector
of the theory is obtained by imposing the divergenceless and
traceless conditions. Some interesting theories are reproduced as
the special cases of the theory and their validity and
successfulness are discussed from group theoretical point of view.
We demonstrated that it is necessary for a theory to be successful,
in dS space-time, if it transforms according to the UIR's of dS
group. We showed that the proposed theory transforms according to
the UIR,s of dS group if the constant coefficients satisfy some
simple conditions. As a result this theory can be used as a
successful model for solving the problems in the framework of
quantum gravity.

As an special case of the theory the linearized Weyl theory of
gravity is reproduced which transforms according to two of the UIR's
of dS group denoted by $\Pi^{\pm}_{2,2}$ and $\Pi^{\pm}_{2,1}$ in
discrete series. We obtained the solution to the conformally
invariant field equation, using the ambient space notations. The
solution can be written as the multiplication of a symmetric rank-2
generalized polarization tensor and a massless minimally coupled
scalar field in dS space. Also we have calculated the conformally
invariant two-point function, in terms of the basic bi-vectors of
the ambient space. It is dS invariant, symmetric and satisfies the
traceless and divergenceless conditions. We therefore claim that the
proposed modified gravity theory under the given restrictions is a
successful one and the introduced procedure can be used as a
theoretical testing for the validity and successfulness of any given
modified theory of gravity.

\setcounter{equation}{0}
\begin{appendix}
\section{Some useful mathematical relations }
The following relations have been used in deriving the linearized
field equations.\b\label {eq:LG 28}
\tilde{R}_{abcd}=H^{2}(\tilde{g}_{ac}\tilde{g}_{bd}-\tilde{g}_{ad}\tilde{g}_{bc}),\e
\b\label {eq:LG 39} \tilde{R}_{ab}=3H^{2}\tilde{g}_{ab},\e \b \label
{eq:LG 40}\tilde{R}=12H^{2},\e
 \b ({\cal{R}}^{c}\;_{dab})_L\equiv\delta
R^{c}\;_{dab}=\frac{1}{2}\left[\nabla_{a}\left(\nabla_{d}h_{b}^{c}
+\nabla_{b}h_{d}^{c}-\nabla^{c}h_{db}
\right)-\nabla_{b}\left(\nabla_{d}h_{a}^{c}
+\nabla_{a}h_{d}^{c}-\nabla^{c}h_{ad} \right) \right].\e \b\label
{eq:LG 32} ({\cal{R}}_{ab})_L\equiv \delta
R_{ab}=\frac{1}{2}\left(\nabla_{a}\nabla_{c}h_{b}^{c}
+\nabla_{b}\nabla_{c}h_{a}^{c}+8H^2h_{ab}-2H^2h'\tilde{g}_{ab}-\Box
h_{ab}-\nabla_{a}\nabla_{b}h'\right).\e \b\label {eq:LG
34}({\cal{R}})_L\equiv\delta R=\nabla_{c}\nabla_{b}h^{cb}-\Box
h'-3H^2h'.\e \b ({\cal{R}}^c_{d})_L\equiv \delta R^c_{d}
=\frac{1}{2}\left(\nabla^{c}\nabla_{a}h_{d}^{a}
+\nabla_d\nabla_{a}h^{ac}+8H^2h^c_{d}-2H^2h'\tilde{g}^c_{d}-\Box
h^c_{d}-\nabla^{c}\nabla_{d}h'\right)-3H^2h^{c}_{d}.\e \b (
{\cal{R}}^{bc})_L\equiv \delta
R^{bc}=\frac{1}{2}\left(\nabla^{c}\nabla_{a}h^{ab}
+\nabla^b\nabla_{a}h^{ac}+8H^2h^{bc}-2H^2h'\tilde{g}^{bc}-\Box
h^{bc}-\nabla^{c}\nabla^{b}h'\right)-6H^2h^{bc}.\e \b (\nabla_a
\nabla_b {\cal{R}})_L\equiv\delta \nabla_a \nabla_b R = \nabla_a
\nabla_b \delta R = \nabla_a \nabla_b \left( \nabla_c\nabla_d
h^{cd}-\Box h'-3H^2 h'\right).\e \b (\Box {\cal{R}})_L\equiv \delta
\Box R =\Box\delta R=\Box \left( \nabla_c\nabla_d h^{cd}-\Box
h'-3H^2 h'\right).\e
 $$ (\nabla_a \nabla_b
{\cal{R}}_{cd})_L\equiv \delta \nabla_a \nabla_b R_{cd}= \frac{1}{2}
\nabla_a \nabla_b \left(\nabla_{c}\nabla_{e}h_{d}^{e}
+\nabla_{d}\nabla_{e}h_{c}^{e}+2H^2h_{cd}-2H^2h'\tilde{g}_{cd}\right.$$\b
\left.-\Box h_{cd}-\nabla_{c}\nabla_{d}h'\right) .\e
$$ (\nabla_a \nabla_b {\cal{R}}^c_{d})_L\equiv \delta \nabla_a \nabla_b
R^c_{d}= \frac{1}{2} \nabla_a \nabla_b
\left(\nabla^{c}\nabla_{e}h_{d}^{e}
+\nabla_{d}\nabla_{e}h^{ce}+2H^2h^c_{d}-2H^2h'\tilde{g}^c_{d}\right.$$\b
\left.-\Box h^c_{d}-\nabla^{c}\nabla_{d}h'\right),\e $$ (\nabla_a
\nabla_b {\cal{R}}^{cd})_L\equiv\delta \nabla_a \nabla_b R^{cd}=
\frac{1}{2} \nabla_a \nabla_b \left(\nabla^{c}\nabla_{e}h^{ed}
+\nabla^{d}\nabla_{e}h^{ce}+2H^2h^{cd}-2H^2h'\tilde{g}^{cd}\right.$$\b
\left.-\Box h^{cd}-\nabla^{c}\nabla^{d}h'\right).\e \b (\Box
{\cal{R}}_{cd})_L\equiv \delta \Box R_{cd}=\frac{1}{2} \Box
\left(\nabla_{c}\nabla_{e}h_{d}^{e}
+\nabla_{d}\nabla_{e}h_{c}^{e}+2H^2h_{cd}-2H^2h'\tilde{g}_{cd}-\Box
h_{cd}-\nabla_{c}\nabla_{d}h'\right).\e

\setcounter{equation}{0}
\section{Details of derivation of Eq.(2.2)}

In this subsection we obtain the field equation through variation of
the action. The action (2.1) can be written as \b I=\frac{1}{16\pi
G}\int d^4
x\sqrt{-g}\left[f({\cal{R}})+b{\cal{R}}^{ab}{\cal{R}}_{ab}
\right],\;\;\;\;\;\;\;\;\;\;\;\;f({\cal{R}})=a_0({\cal{R}}-2\Lambda
) + a{\cal{R}}^2,\e   \b \delta
\left[\sqrt{-g}f({\cal{R}})\right]=f({\cal{R}})\delta
\sqrt{-g}+\sqrt{-g}f'({\cal{R}})\delta{\cal{R}},\e

\b \delta \sqrt{-g}=-\frac{1}{2}\sqrt{-g}g_{ab}\delta g^{ab},\e \b
\delta{\cal{R}}=\delta\left(g^{ab}{\cal{R}}_{ab}\right)={\cal{R}}_{ab}\delta
g^{ab}+g^{ab}\delta{\cal{R}}_{ab},\e \b
\delta{\cal{R}}_{ab}=\nabla_c\delta\Gamma^c_{ab}-\nabla_b\delta\Gamma^c_{ac}.\e
The statement in Eq.(B.5) is the difference of two connections, it
transforms as a tensor. one can show that \b
\delta\Gamma^c_{ab}=\frac{1}{2}g^{cd}\left(\nabla_a \delta
g_{bd}+\nabla_b \delta g_{ad}-\nabla_d \delta g_{ab} \right),\e and
substituting in Eq.(B.5)we have  \b
\delta{\cal{R}}_{ab}=\frac{1}{2}\left(\nabla^c\nabla_a \delta
g_{bc}+\nabla^c\nabla_b \delta g_{ac} -\nabla_a \nabla_bg^{cd}\delta
g_{cd}- \Box \delta g_{ab}\right).\e Now return to Eq.(B.4) we have
\b \delta{\cal{R}}={\cal{R}}_{ab}\delta g^{ab}+g_{ab}\Box \delta
g^{ab}-\nabla_a\nabla_b \delta g^{ab}.\e Using Eq.(B.3) and Eq.(B.8)
in Eq.(B.2) we have \b \delta
\left[\sqrt{-g}f({\cal{R}})\right]=\sqrt{-g}\left[f'({\cal{R}}){\cal{R}}_{ab}-\frac{1}{2}g_{ab}f({\cal{R}})+f'({\cal{R}})\left(
g_{ab}\Box -\nabla_a \nabla_b \right)\right]\delta g^{ab}. \e \b
\delta \left( {\cal{R}}^{ab} {\cal{R}}_{ab}\right)=\delta \left(
g^{ca}g^{bd}{\cal{R}}_{cd}
{\cal{R}}_{ab}\right)=2\left({\cal{R}}^a_{c} {\cal{R}}_{ab}\delta
g^{bc}+{\cal{R}}^{ab} \delta {\cal{R}}_{ab}\right),\e and noting
Eq.(B.7) we can show \b \delta \left( {\cal{R}}^{ab}
{\cal{R}}_{ab}\right)=2{\cal{R}}^a_{c} {\cal{R}}_{ab}\delta
g^{bc}+{\cal{R}}^{ab}\left(\nabla^c\nabla_a \delta
g_{bc}+\nabla^c\nabla_b \delta g_{ac} -\nabla_a \nabla_bg^{cd}\delta
g_{cd}- \Box \delta g_{ab}\right).\e Now it is easy to show that $$
\delta \left(\sqrt{-g} {\cal{R}}^{ab}
{\cal{R}}_{ab}\right)=\sqrt{-g}\left[-\frac{1}{2}g_{ab}{\cal{R}}_{cd}{\cal{R}}^{cd}+2{\cal{R}}^c_{a}
{\cal{R}}_{cb}- {\cal{R}}^{c}_a\nabla_b\nabla_c\right.$$ \b \left.-
{\cal{R}}^{c}_b\nabla_a\nabla_c+{\cal{R}}^{cd}\nabla_c\nabla_d
g_{ab}+ {\cal{R}}_{ab}\Box \right]\delta g^{ab}. \e

$$\delta I=\frac{1}{16\pi G}\int d^4 x\sqrt{-g}\left[f'({\cal{R}}){\cal{R}}_{ab}-\frac{1}{2}g_{ab}f({\cal{R}})
-\frac{b}{2}g_{ab}{\cal{R}}_{cd}{\cal{R}}^{cd}+2b{\cal{R}}^c_{a}
{\cal{R}}_{cb} \right.$$ \b \left.+f'({\cal{R}}) g_{ab}\Box
-f'({\cal{R}})\nabla_a \nabla_b
 -b{\cal{R}}^{c}_a\nabla_b\nabla_c -b
{\cal{R}}^{c}_b\nabla_a\nabla_c+b{\cal{R}}^{cd}\nabla_c\nabla_d
g_{ab}+ b{\cal{R}}_{ab}\Box \right]\delta g^{ab}. \e

Doing integration by part on the last six terms two times leads to
$$\delta I=\frac{1}{16\pi G}\int d^4 x\sqrt{-g}\left[f'({\cal{R}}){\cal{R}}_{ab}-\frac{1}{2}g_{ab}f({\cal{R}})
-\frac{b}{2}g_{ab}{\cal{R}}_{cd}{\cal{R}}^{cd}+2b{\cal{R}}^c_{a}
{\cal{R}}_{cb} \right.$$ \b \left.+g_{ab}\Box f'({\cal{R}})
-\nabla_a \nabla_bf'({\cal{R}})
 -b\nabla_c\nabla_b{\cal{R}}^{c}_a -b\nabla_c\nabla_a
{\cal{R}}^{c}_b+b\nabla_c\nabla_d{\cal{R}}^{cd} g_{ab}+
b\Box{\cal{R}}_{ab} \right]\delta g^{ab}. \e Noting that action
remains invariant under variation of the metric and putting $\delta
I=0, $ results in Eq.(2.2).

\setcounter{equation}{0}
\section{Details of derivation of Eq.(2.7)}

Now $\H^{(0)}_{ab}$ can be written as
$$\H^{(0)}_{ab}=\tilde{R}_{ab}+\delta
R_{ab}-\frac{1}{2}(\tilde{R}+\delta
R)(\tilde{g}_{ab}+h_{ab})+\Lambda (\tilde{g}_{ab}+h_{ab})$$ \b=
\tilde{H}^{(0)}_{ab}+\delta
R_{ab}-\frac{1}{2}(\tilde{R}h_{ab}+\tilde{g}_{ab}\delta R )+\Lambda
h_{ab}+ \mbox{nonlinear terms}.\e Using Eqs.(A.3), (A.5) and (A.6)
in (C.1) leads to the expression presented in Eq.(2.7).

\setcounter{equation}{0}
\section{Details of derivation of Eq.(2.9)}
Now, $\H^{(1)}_{ab}$ can be written as
$$ \H_{ab}^{(1)}+2(\tilde{R}+\delta
R)(\tilde{R}_{ab}+\delta R_{ab})-2(\nabla_a \nabla_b
\tilde{R}+\delta \nabla_a \nabla_b R)$$
\b-\frac{1}{2}(\tilde{g}_{ab}+h_{ab})\left[(\tilde{R}^2+2\tilde{R}\delta
R)-4(\Box \tilde{R}+\delta\Box R)\right],\e which may be written as
can be written as  $$
\H_{ab}^{(1)}=\tilde{H}_{ab}^{(1)}-2\left(\delta \nabla_a \nabla_b
R-\tilde{R}\delta R_{ab}-\tilde{R}_{ab}\delta
R\right)-\tilde{g}_{ab}(\tilde{R} \delta R-2\delta\Box R)-
\frac{1}{2}h_{ab}\tilde{R}^2+\mbox{nonlinear terms}.$$ Making use of
Eqs. (A.5), (A.6), (A.9) and (A.10) we have $\H^{(1)}_{ab}=
\tilde{H}^{(1)}_{ab}+H^{(1)}_{ab},$ and $H^{(1)}_{ab}$ is the same
as given in Eq.(2.9).

\setcounter{equation}{0}
\section{Details of derivation of Eq.(2.11)}
Eq.(2.4) can be written in the following form
$$ \H_{ab}^{(2)}=\left(\Box \tilde{R}_{ab}+\delta
\Box R_{ab} \right)-\left( \nabla_c \nabla_a \tilde{R}^c_b+\delta
\nabla_c \nabla_a R^c_b \right)-\left(\nabla_c \nabla_b
 \tilde{R}^c_a+\delta \nabla_c \nabla_b R^c_a\right) $$ $$ -\frac{1}{2}(\tilde{g}_{ab}+h_{ab})\left[(\tilde{R}^{cd}+\delta
R^{cd})(\tilde{R}_{cd}+\delta R_{cd})-2(\nabla_c \nabla_d
\tilde{R}^{cd}+\delta \nabla_c \nabla_d R^{cd})\right]
$$\b+2\left(\tilde{R}_{a}^c+\delta
R_{a}^c\right)\left(\tilde{R}_{cb}+\delta R_{cb}\right).\e It can be
written as  $$ \H_{ab}^{(2)}=\tilde{H}_{ab}^{(2)}+\delta\Box
R_{ab}-\delta\nabla_c\nabla_aR^c_b-\delta\nabla_c\nabla_bR^c_a+2(\tilde{R}^c_a\delta
R_{cb}+\tilde{R}_{cb}\delta R_{a}^c)$$
\b-\frac{1}{2}\tilde{g}_{ab}(\tilde{R}^{cd}\delta
R_{cd}+\tilde{R}_{cd}\delta R^{cd}-2\delta \nabla_c \nabla_d
R^{cd})-\frac{1}{2}h_{ab}\tilde{R}^{cd}\tilde{R}_{cd}+
\mbox{nonlinear terms}.\e

Now using the relations given in appendix-A we can show that \b
\tilde{R}^c_a\delta R_{cb}=
\frac{3}{2}H^2\left(\nabla_{a}\nabla_{c}h_{b}^{c}
+\nabla_{b}\nabla_{c}h_{a}^{c}+8H^2h_{ab}-2H^2h'\tilde{g}_{ab}-\Box
h_{ab}-\nabla_{a}\nabla_{b}h'\right),\e \b \tilde{R}_{cb}\delta
R_{a}^c =\frac{3}{2}H^2 \left(\nabla_{a}\nabla_{c}h_{b}^{c}
+\nabla_{b}\nabla_{c}h_{a}^{c}+2H^2h_{ab}-2H^2h'\tilde{g}_{ab}-\Box
h_{ab}-\nabla_{a}\nabla_{b}h'\right) ,\e \b \delta \nabla_c \nabla_d
R^{cd}= \nabla_c \nabla_d(\nabla^c \nabla_f h^{fd}+\nabla^d \nabla_f
h^{fc})+2H^2\nabla_c \nabla_d h^{cd}-2H^2\Box h'- \nabla_c \nabla_d
\Box h^{cd}-\nabla_c \Box \nabla^ch', \e \b \tilde{R}^{cd}\delta
R_{cd}=3H^2\left(\nabla_c \nabla_d h^{cd}-\Box h'\right) ,\e
 \b\tilde{R}_{cd}\delta R^{cd}=3H^2\left(\nabla_c \nabla_dh^{cd}-\Box h'-6H^2 h' \right).\e

Substituting in Eq.(E.2) we obtain
$\H_{ab}^{(2)}=\tilde{H}_{ab}^{(2)}+H_{ab}^{(2)}$ and $H_{ab}^{(2)}$
is the statement given in Eq.(2.11).

 In derivation steps, the following identities have been used $$\nabla_c\nabla_a\Box
h^c_b=\Box\nabla_a\nabla_ch^c_b+2H^2\nabla_{a}\nabla_{c}h_{b}^{c}-2H^2\nabla_{b}\nabla_{c}h_{a}^{c}
+2H^2\tilde{g}_{ab}\nabla_c \nabla_d
h^{cd}$$\b-2H^2\nabla_a\nabla_bh'+4H^2\Box
h_{ab}-H^2\tilde{g}_{ab}\Box h',\e \b
\nabla_c\nabla_a\nabla^c\nabla_dh^d_b=\Box\nabla_a\nabla_d
h^d_b-H^2\nabla_b\nabla_dh^d_c+H^2\tilde{g}_{ab}\nabla_c \nabla_d
h^{cd},\e \b \nabla_c\nabla_a\nabla_b\nabla^ch'=\Box\nabla_a\nabla_b
h'-H^2\nabla_a\nabla_bh'+H^2\tilde{g}_{ab}\Box h',\e \b \nabla_c
\nabla_a \nabla_b \nabla_d h^{cd}=\nabla_a\nabla_b \nabla_c \nabla_d
h^{cd}
+4H^2\nabla_a\nabla_ch^c_b+3H^2\nabla_b\nabla_ch^c_a-H^2\tilde{g}_{ab}\nabla_c
\nabla_d h^{cd},\e \b
\nabla_c\nabla_a\nabla_b\nabla_dh^{cd}=\nabla_a\nabla_b \nabla_c
\nabla_d
h^{cd}+4H^2\nabla_a\nabla_ch^c_b+3H^2\nabla_b\nabla_ch^c_a-H^2\tilde{g}_{ab}\nabla_c
\nabla_d h^{cd},\e \b\Box \nabla_a \nabla_b h'= \nabla_a\nabla_b
\Box h' +8H^2\nabla_a\nabla_bh'-2H^2\tilde{g}_{ab}\Box h'.\e

\setcounter{equation}{0}
\section{Details of derivation of Eq.(3.3)}
By imposing the tensor field  \b \K=\theta\phi_1+ {\cal {S}}\bar
{Z}_1K+D_2K_g,\e to obey the field equation \b
(Q^{(1)}_2+4)(Q^{(1)}_2+6)\K=0 ,\e and making use of the following
identities \cite{gagata}, \b Q^{(1)}_2(\theta\phi)=\theta  Q^{(1)}_0
\phi ,\e \b Q^{(1)}_2D_2K_g=D_2Q^{(1)}_1K_g ,\e
 \b Q^{(1)}_2{\cal S}( \bar{Z} K) ={\cal S}\left[\bar{Z}(Q^{(1)}_1-4)K\right] -2H^2D_2(x.Z)K+4\theta Z.K,\e
 we have
$$ (Q^{(1)}_2+4)\left(\left[(Q^{(1)}_0+6)\phi_1+4Z_1.K
\right]\theta\right.$$\b\left.+{\cal
S}\left[\bar{Z}_1(Q^{(1)}_1+2)K\right]+D_2\left[(Q^{(1)}_1+6)K_g-2H^2(x.Z_1)K
\right] \right)=0.\e Making use of Eqs.(F.3)-(F.5) in Eq.(F.6) once
again, we obtain
$$\theta
\left[(Q^{(1)}_0+4)(Q^{(1)}_0+6)\phi_1+4Q^{(1)}_0Z_1.K+4(Q^{(1)}_1+2)x.Z_1K+16
Z_1.K \right]+{\cal
S}\bar{Z}_1\left[Q^{(1)}_1(Q^{(1)}_1+2)K\right]$$ \b
+D_2\left[(Q^{(1)}_1+4)(Q^{(1)}_1+6)K_g-2H^2(x.Z_1)(Q^{(1)}_1+6)K-2H^2Q^{(1)}_1(x.Z_1K)
\right]=0 .\e Using the conditions $x.K=0=\partial.K$ in Eq.(2.29)
we have \b (Q^{(1)}_1+2)K=Q^{(1)}_0K,\e from which we can write \b
Q^{(1)}_0(Z_1.K)=(Q^{(1)}_1+2)Z_1.K. \e Substituting (F.9) in
Eq.(F.7) results in
$$\theta
\left[(Q^{(1)}_0+4)(Q^{(1)}_0+6)\phi_1+8Q^{(1)}_0Z_1.K+16 Z_1.K
\right]+{\cal S}\bar{Z}_1\left[Q^{(1)}_1(Q^{(1)}_1+2)K\right]$$ \b
+D_2\left[(Q^{(1)}_1+4)(Q^{(1)}_1+6)K_g-2H^2(x.Z_1)(Q^{(1)}_1+6)K-2H^2Q^{(1)}_1(x.Z_1K)
\right]=0 .\e It is easy to show that \b Q^{(1)}_1(x.Z_1K)= x.Z_1(
Q^{(1)}_1-4)K-2Z_1.D_1 K +2xZ_1.K.\e Now combining Eqs.(F.10) and
(F.11) leads to the following equation $$\theta
\left[(Q^{(1)}_0+4)(Q^{(1)}_0+6)\phi_1+8(Q^{(1)}_0+2)Z_1.K\right]+{\cal
S}\bar{Z}_1\left[Q^{(1)}_1(Q^{(1)}_1+2)K\right]$$ \b
+D_2\left[(Q^{(1)}_1+4)(Q^{(1)}_1+6)K_g-4H^2\left((Q^{(1)}_1+5)(x.Z_1K)+Z_1.D_1K-xZ_1.K\right)
\right]=0,\e which results in Eq.(3.3).

 \end{appendix}

\end{document}